\documentclass[conference]{IEEEtran}
\IEEEoverridecommandlockouts
\usepackage{cite}
\usepackage{amsmath,amssymb,amsfonts}
\usepackage{algorithmic}
\usepackage{graphicx}
\usepackage{textcomp}
\usepackage{xcolor}
\usepackage{caption}
\usepackage{subcaption}
\usepackage[ruled]{algorithm2e}
\usepackage{caption}
\usepackage{subcaption}
\usepackage{epstopdf}
\usepackage{multirow}
\usepackage{svg}
\usepackage{balance}

\newcommand{\ps}{\mathbf{p}_s}

\newcommand{\p}{\mathbf{p}}

\newcommand{\dt}{\text{d}t}

\newtheorem{assumption}{\textbf{Assumption}}
\newtheorem{theorem}{\textbf{Theorem}}
\newtheorem{definition}{Definition}
\newtheorem{corollary}{Corollary}

\newcommand{\TT}{\mathsf{T}}
\newcommand{\Lcal}{\mathcal{L}}
\newcommand{\Acal}{\mathcal{A}}

\renewcommand{\Vec}[1]{\mathbf{#1}}
\newcommand{\param}{\mathbf{w}}

\def\BibTeX{{\rm B\kern-.05em{\sc i\kern-.025em b}\kern-.08em
    T\kern-.1667em\lower.7ex\hbox{E}\kern-.125emX}}
\begin{document}

\title{Mismatch-Robust Underwater Acoustic Localization Using A Differentiable Modular Forward Model
\thanks{This work has been supported by the Office of Naval Research (ONR) under grant N00014-19-1-2662.}}

\author{\IEEEauthorblockN{Dariush Kari}
\IEEEauthorblockA{
\textit{University of Illinois Urbana-Champaign}\\
Urbana, IL, USA \\
dkari2@illinois.edu}
\and 
\IEEEauthorblockN{Yongjie Zhuang}
\IEEEauthorblockA{
\textit{Stony Brook University}\\
Stony Brook, NY, USA \\
yongjie.zg@gmail.com}
\and
\IEEEauthorblockN{Andrew C. Singer}
\IEEEauthorblockA{
\textit{Stony Brook University}\\
Stony Brook, NY, USA \\
andrew.c.singer@stonybrook.edu}
}

\maketitle

\begin{abstract}
In this paper, we study the underwater acoustic localization in the presence of environmental mismatch. Especially, we exploit a pre-trained neural network for the acoustic wave propagation in a gradient-based optimization framework to estimate the source location. To alleviate the effect of mismatch between the training data and the test data, we simultaneously optimize over the network weights at the inference time, and provide conditions under which this method is effective. Moreover, we introduce a physics-inspired modularity in the forward model that enables us to learn the path lengths of the multipath structure in an end-to-end training manner without access to the specific path labels. We investigate the validity of the assumptions in a simple yet illustrative environment model.
\end{abstract}


\section{Introduction}
Underwater acoustic (UWA) localization can be performed using physics-based or learning-based approaches. Physics-based methods usually rely on prior knowledge about the environment such as the sound speed profile and bathymetry \cite{baggeroer1988matched}, which is rarely available in practice. On the contrary, learning-based methods \cite{niu2023advances, niu2017source, yangzhou2019deep, chen2021model, wang2018underwater} do not require a priori knowledge about the environment but usually require a significant amount of training data to implicitly learn the environment and become operational, which is (usually prohibitively) expensive. As a result, there is a need to build more data-efficient learning-based algorithms for the UWA localization task. This paper seeks a model that can be trained on an environment and then be adapted to a test environment with access to few unlabeled data and no access to the training data.\par

UWA localization using deep learning algorithms often generalizes poorly to new environments \cite{liu2023unsupervised}. Strategies to address this issue include transfer learning\cite{ge2022label, wang2019deep,yosinski2014transferable}, data augmentation \cite{liu2021deep,castro2021impact}, and domain adaptation \cite{liu2023unsupervised, long2023deep}. Transfer learning involves training a deep learning model on a synthetic dataset, e.g., a dataset created using UWA propagation models like Bellhop \cite{porter2011bellhop} or Kraken \cite{porter1992kraken}, and fine-tuning parts of the model to the small labeled dataset available from the test environment for real-world applications. Alternatively, one might augment those labeled test data to create a larger dataset that can be used for training. However, the data-deprived and highly non-stationary nature of underwater environments hinders the practicality of such approaches. Nevertheless, a promising alternative is the test-time adaptation \cite{wang2020tent}, where the model will be adapted to a new environment without any supervised training over the samples from the test environment.\par

Test-time domain adaptation \cite{wang2020tent} can be interpreted as the procedure of adapting a pre-trained model to a test environment using some prior belief about the test data distribution or a prior knowledge about the features of the data that can be obtained in an unsupervised manner. For instance, inverse models that take the received signal and output the source location, can be adapted to a new environment by implicitly inferring the source characteristics or imposing constraints on the outputs to follow an a priori distribution \cite{kari2024joint}. Nevertheless, such methods usually require a significant amount of unlabeled data, which is not always available. An alternative approach can leverage physics-based priors \cite{li2023data} incorporated into a forward model (which is not straightforward to be incorporated in inverse models) that maps the source location to the received signal. The localization can then be performed by optimization over the inputs of the forward model \cite{kari2023gradient, chitre2023differentiable}. To exploit this characteristic of forward modeling, we propose a \emph{modular} network based on acoustic ray propagation \cite{jensen2011computational}.\par

Off-the-shelf neural network architectures are capable of end-to-end training, at the expense of poor transferability, interpretability, and ignoring the multipath structure. On the other hand, training directly over the path parameters such as the time/amplitude-of-arrivals instead of an end-to-end training is usually a cumbersome task as it needs path labeling. Therefore, we propose a modular structure that respects physics and supports end-to-end training. Modularity increases the interpretability of the network as each module in the network is responsible for a specific task. In addition, it leverages physics because the interaction between different modules is based on physics and the network does not need to learn them. These features enable us to train the model in an end-to-end fashion, without the need for labeled paths. Furthermore, to use the model in a new environment, we do not need to retrain (adapt) the whole model; it is sufficient to retrain only the modules that are known to be different in the new environment.\par

There are optimization-based solutions for inverse problems \cite{arridge2019solving}, which can leverage the merits of gradient-based (GB) optimization whenever the forward model is differentiable. Our forward model is a deep learning-based structure and thus differentiable. With GB optimization over the input, we can address the UWA localization problem \cite{kari2022underwater}. Nevertheless, to mitigate performance degradation caused by mismatches between the training and testing environments, we propose optimizing not only over the inputs but also over the network parameters, effectively enabling test time adaptation.\par

\textbf{Notation}: Bold letters denote vectors or matrices. A subscript $s$ or $r$, respectively, refers to the source or receiver. The $i$-th entry of a vector $\Vec{a}$ is denoted by $[\Vec{a}]_i$. The vector $\Vec{p}$ denotes a Cartesian location. Moreover, all-one vector is denoted by $\Vec{1}$.

\section{Gradient-Based Localization (GBL)} \label{sec:GBL}

Localization methods relying solely on times-of-arrival (TOA) are not always optimal \cite{kari2023gradient}. Therefore, we propose a maximum likelihood estimator (MLE; under the white Gaussian noise assumption) for determining the location directly, rather than the TOAs. However, due to the lack of a closed-form solution, numerical optimization methods are necessary. To mitigate the high computational costs associated with global optimization techniques, we exploit GB optimization \cite{kari2023gradient}. Nonetheless, to initialize the GBL, we use the TOA-based method of \cite{kari2023gradient} as a form of convex relaxation \cite{naddafzadeh2013second, papalia2023score}. GBL requires a differentiable model for acoustic wave propagation and is appropriate when the gradient vanishes at the loss minimum \cite{kari2023gradient}. In this paper, we use a modular model $f_{\param}$, which includes a parametric representation of the source signal and a neural network for the geometry extraction (parameterized by $\param  \in \mathbb{R}^{n_{\param}}$), all of which are differentiable. Assuming additive white Gaussian noise (AWGN), the maximum likelihood estimator for the source location is

\begin{equation}\label{eq:GBL}
    \hat \p = \arg \min_{\p} \; \; \int_0^T (r(t) - f_{\param}(t,\p) )^2 \dt,
\end{equation}
where $r(t)$ is the received signal and the $f_{\param}(t, \p)$ is the model output for the hypothetical source location $\p \in \mathbb{R}^{n_{\p}}$, and $T$ is the observation length. 

\section{Modular Network} \label{sec:modular}
\begin{figure}[ht]
\centering
    \includegraphics[width= \linewidth]{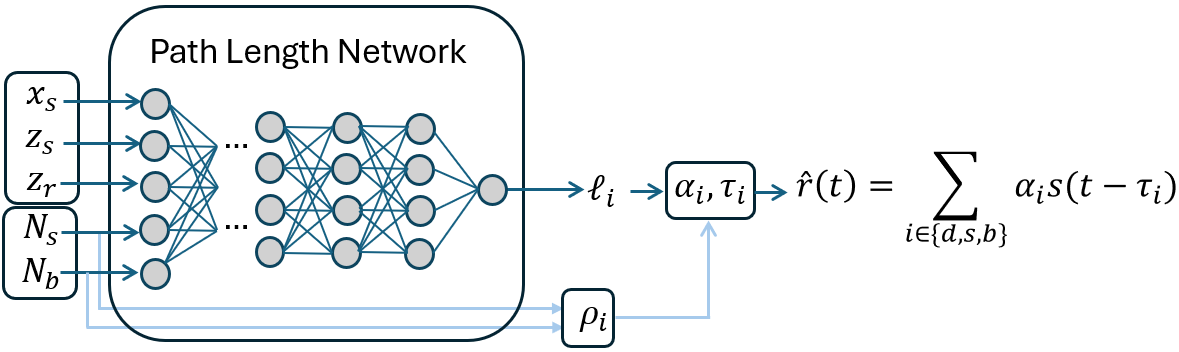}
    \caption{The module for path length computation is a fully connected network. Each path $i$ is identified by its number of surface and bottom reflections, $N_s$ and $N_b$. Also, $\rho_i$ indicates the reflection coefficient experienced during the $i$-th path. Here, $\rho_i = (-1)^{N_s(i)}$}
    \label{fig:modular}
\end{figure}

To leverage the known ray propagation laws in our problem, we use a hybrid model as shown in Fig. \ref{fig:modular}, where the Path Length Network (PLN) is a fully connected neural network. The network outputs the path lengths based on source and receiver locations and the number of surface and bottom reflections, denoted respectively by $N_s$ and $N_b$. However, the training is implicit, i.e., the network seeks to minimize the squared error between the observed received signal $r(t)$ and the reconstructed signal $\hat r(t)$, as opposed to being directly supervised with the path lengths. However, since the network is modular, this procedure yields learning of the multipath structure.\par

The modules in our model include a source signal module that assumes a known waveform $s(t)$, a sound speed parameter $c$, the PLN, and a module that converts the path length $\ell_i$ to $\alpha_i$ and $\tau_i$. Note that if the signal is not known, one may use a parametric form of it to be optimized or another network for its implicit representation \cite{sitzmann2020implicit}. Similarly, our modular structure allows us to optimize over the sound speed as well as the PLN weights.\par 

The input to the PLN consists of $5$ scalars: 2 source coordinates $x_s$ and $z_s$ and one receiver coordinate $z_r$ (note that we assume $x_r = 0$ as the reference), and $2$ integer variables $N_s$ and $N_b$, which indicate the number of surface reflections and bottom reflections along the path, respectively. Note that in the three-ray environment that we use, these variables can distinguish different paths, however, for more complex scenarios that involve higher order reflections, one may use another representation of the paths (e.g., the departure angle of each path as used in Bellhop \cite{porter2011bellhop}).\par

The training loss function is defined as
\begin{align}
    \Lcal_{\text{tr}}(\param) &= \frac 1 {N_{tr}}\sum_{k=1}^{N_{tr}} \int_0^T \left( r^{(k)}(t) - f_{\param}(t, \p^{(k)}) \right)^2 \dt,
\end{align}
where $N_{tr}$ is the size of the training set and the training dataset is $\{(\p^{(k)}, r^{(k)} (.))\}_{k=1}^{N_{tr}}$. The pre-trained model is then
\begin{equation}
    \param^{\mathsf{tr}} = \arg \min_\param \Lcal_{\text{tr}}(\param).
\end{equation}

Note that we do not use a regularization term here, since the physics-based modular network provides a ray-based solution to the Helmholtz equation \cite{jensen2011computational}. As shown in Fig. \ref{fig:modular}, the network learns to estimate the path lengths $\ell_i$, which are ultimately combined with physical priors (here in the form of $\rho_i$) to make $\alpha_i$ and $\tau_i$ consistent with the underlying physics.

\section{Adapting To The Environment}

\begin{figure}[ht]
\centering
    \includegraphics[width= 0.5 \linewidth]{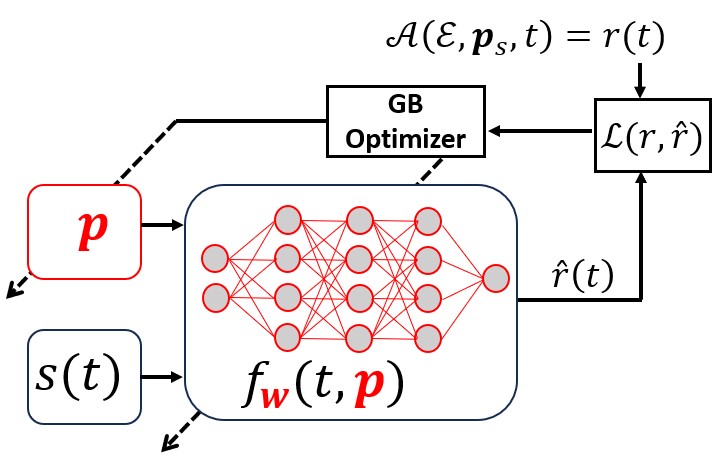}
    \caption{Block diagram of forward domain adaptation. During the adaptation, the loss minimization is performed with respect to both $\p$ and $\param$.}
    \label{fig:forward_da_block}
\end{figure}

To overcome the challenges posed by mismatches between the training and testing environments, as shown in Fig. \ref{fig:forward_da_block}, we optimize simultaneously over the model input (source location) and the model parameters. In this section, we show  sufficient conditions (Assumptions \ref{assm:represent}-\ref{assm:continuity}) under which the proposed adaptation method achieves a bounded error for localization in the mismatched environment. For representation simplicity, we assume a known and fixed source location in this section.

\begin{assumption}[Nature Representability]\label{assm:represent} \hfill \\
The received signal corresponding to the source location $\ps \in \mathbb{R}^{n_p}$ can be represented as $r(t) = \Acal(\mathbf{e}, \ps, t)$, where $\mathbf{e} \in \mathbb{R}^{n_e}$ denotes the environmental parameters (e.g., sound speed profile and bathymetry, among other parameters), and $\Acal: \mathbb{R}^{n_e + n_p + 1} \to \mathbb{R}$ is a scalar function of a multivariate input.
\end{assumption}

To account for the environmental mismatch, instead of \eqref{eq:GBL}, we solve $(\hat \param, \hat \p) = \arg \min_{(\param, \p)} \Lcal_{\text{DA}}(\mathbf{e}, \ps, \param, \p)$, where
\begin{align}
    \Lcal_{\text{DA}}(\mathbf{e}, \ps, \param, \p) &= \int_0^T (r(t) - f_{\param}(t,\p) )^2 \dt + \frac \gamma 2 \|\param - \param_{\mathsf{tr}}\|^2,
\end{align}
and $\param_{\mathsf{tr}} \in \mathbb{R}^{n_w}$ is the pre-trained model parameters,$\gamma \geq 0$ is a regularization hyper-parameter determined based on our prior belief about the amount of mismatch between the training and the test environments, and $\Lcal_{DA}: \mathbb{R}^{n_w+n_e+2n_p} \to \mathbb{R}^{\geq 0}$ is the domain adaptation loss. Our proposed Domain Adaptive GBL (DA-GBL) algorithm takes the parameter $\param_{\mathsf{tr}}$, $\gamma$, $r(t)$ for $t \in [0, T]$, and an initial estimate $\hat{\p}_0$ and generates an estimate $\hat{\p}$ using a GB optimization over the loss $\Lcal_{DA}$. We denote this algorithm concisely by $\hat{\Vec{v}} = \Gamma(\param_{\mathsf{tr}}, \gamma, \hat{\p}_0, r(.))$, where $\hat{\Vec{v}} = [\hat{\param}^\TT, \hat{\p}^\TT]^\TT$, and $r(.)$ denotes the received signal $r(t)$ for $t \in [0, T]$. The initial estimate $\hat{\p}_0$ can be obtained from a TOA method based on a simple model of the environment (which is not necessary an accurate model of the environment). If the model has been trained using the data corresponding to the environment $\mathbf{e}_{\mathsf{tr}}$, but it is used for localization in the environment $\mathbf{e}_{\mathsf{test}} = \mathbf{e}_{\mathsf{tr}} + \boldsymbol{\epsilon}$, we want to bound the following term
\begin{equation}\label{eq:Gamma_perturb}
\|  \Gamma(\param_{\mathsf{tr}}, \gamma, \hat{\p}_0, \Acal(\mathbf{e}_{\mathsf{tr}}, \ps, .)) -  \Gamma(\param_{\mathsf{tr}}, \gamma, \hat{\p}_0, \Acal(\mathbf{e}_{\mathsf{test}}, \ps, .)) \|.
\end{equation}

Since the output of $\Gamma$ is a stationary point of the gradient of $\Lcal_{DA}$, it suffices that we bound the perturbations of the stationary points of the gradient when there is a perturbation $\boldsymbol{\epsilon}$ in $\mathbf{e}$. The gradient of $\Lcal_{DA}$ with respect to $\Vec{v} = [\param^\TT, \p^\TT]^\TT$ is a function $G:\mathbb{R}^{n_w+n_e+2n_p} \to \mathbb{R}^{n_w+n_p}$ defined as
\begin{equation}
    G(\Vec{v}, \mathbf{e}, \ps) = \nabla_{\Vec{v}} \Lcal_{DA} (\mathbf{e}, \ps, \Vec{v}).
\end{equation}

\emph{Note}: In the AWGN case, if $f_{\param}$ faithfully represents the noiseless received signal, then $\text{Pr}(r(t)|\mathbf{e}, \ps)$ is Gaussian with the mean $f_{\param}$, hence, $G$ becomes proportional to the Fisher score function \cite{moulin2018statistical}. The assumption in that case would be that the forward model can exactly recover the noiseless signal.\par

The following assumptions provide sufficient conditions for $G$ and $\Lcal_{DA}$ so that we can bound \eqref{eq:Gamma_perturb}.

\begin{definition}
    A cube $C_\sigma(\Vec{a})$ is an $\ell_\infty$-neighborhood of $\Vec{a} \in \mathbb{R}^{n_a}$ defined by 
    \[
    C_{\sigma}(\Vec{a}) = \{\Vec{b}: \; \vline [\Vec{b} - \Vec{a}]_i \; \vline \leq \sigma,  \; \forall i \in \{1,2, ..., n_a \} \}.
    \]
\end{definition}

\begin{assumption}[$L$-Lipschitz gradient $G$] \label{assm:Lipschitz}\hfill \\
$G$ is assumed to be $L$-Lipschitz \cite{shalev2014understanding} with respect to $\mathbf{e}$, i.e.,
\begin{equation}
    \| G(\Vec{v}, \mathbf{e} + \boldsymbol{\epsilon}, \ps) - G(\Vec{v}, \mathbf{e}, \ps) \| \leq L \| \boldsymbol{\epsilon} \|.
\end{equation}

\end{assumption}

\begin{assumption}[$\lambda$-Strong Convexity of $\Lcal$] \label{assm:convexity}\hfill \\

$\Lcal_{DA}$ is a $\lambda$-strongly convex \cite{shalev2014understanding} function of $\Vec{v} = [\param^\TT, \p^\TT]^\TT$ on a cube $C_{\sigma}$ around $\Vec{v}_0$. Therefore,
\begin{multline}
    \Lcal_{DA}(\mathbf{e}, \ps, \Vec{v}) \geq \Lcal_{DA}(\mathbf{e}, \ps, \Vec{v}_0) + \\ \nabla_{\Vec{v}} \Lcal_{DA}(\mathbf{e}, \ps, \Vec{v}_0)^{\mathsf{T}} (\Vec{v} - \Vec{v}_0) + \frac{\lambda}{2} \|\Vec{v} - \Vec{v}_0\|^2.
\end{multline}

\end{assumption}

It is useful to define a parameterized directional gradient vector $\mathbf{g}: \mathbb{R} \to \mathbb{R}^{n_w+n_p}$ by
\begin{equation}
    \mathbf{g}(\kappa) = G(\Vec{v}_0 + \kappa \Vec{H}(\Vec{v}_0)^{-1}\Vec{1}, \mathbf{e}_{\mathsf{tr}}, \ps),
\end{equation}
where $\Vec{v}_0$ is a root of $G$ that is the output of $\Gamma$ and we define $\Vec{H}(\Vec{v}) \triangleq \nabla_{\Vec{v}} G = \nabla^2_{\Vec{v}} \Lcal_{DA}(\mathbf{e}, \ps,\Vec{v})$ as the Hessian of the loss. Because of the $\lambda$-strong convexity, $\Vec{H}(\Vec{v}) \geq \lambda \mathbf{I}$, hence, $\Vec{H}(\Vec{v})^{-1}$ exists. Observe that $\mathbf{g}(0) = G(\Vec{v}_0, \mathbf{e}_{\mathsf{tr}}, \ps) = \Vec{0}$ and $\mathbf{g}'(0) = \Vec{H}(\Vec{v}_0) \Vec{H}(\Vec{v}_0)^{-1}\Vec{1} = \Vec{1}$.

\begin{assumption}[Continuity and Twice Differentiability of $G$]\label{assm:continuity}\hfill \\
The loss gradient $G$ is continuous and differentiable over its domain, and $\mathbf{g}$ is twice differentiable. This is a valid assumption as we have control over the model $f_\param$. As a result, $\mathbf{g}$ can be expanded by Taylor series as
\[
\mathbf{g}(\kappa) = \mathbf{g}'(0) \kappa + \frac{\mathbf{g}''(z)} 2 \kappa^2,
\]
for any $\kappa \in [0, \sigma]$ and some $z \in [0, \kappa]$, where $\Vec{g}'$ and $\Vec{g}''$ denote the first- and second-order derivatives of $\mathbf{g}(\kappa)$ with respect to $\kappa$. We also assume that $\mathbf{g}''(z)$ is bounded for any $z \in [0, \sigma]$.

\end{assumption}

Let us define the element-wise supremum $\xi_i$ as $\xi_i = \max_{z \in [0, \sigma]} \; | [\mathbf{g}''(z)]_i |$, and $\xi = \max_{i} \xi_i$. From these definitions, we see that 
\begin{equation}\label{eq:xi_bound}
\frac{ \vline \; [\mathbf{g}''(z)]_i \; \vline}{\xi} \leq 1.
\end{equation}

\begin{theorem} [Perturbation in The Solution $\Vec{v}_0$] \label{thm:PLSP}
    Suppose that $\Vec{v}_0 = \Gamma(\param_{\mathsf{tr}}, \gamma, \hat{\p}_0, \Acal(\mathbf{e}_{\mathsf{tr}}, \ps, .))$. If the Assumptions \ref{assm:represent}, \ref{assm:Lipschitz}, \ref{assm:convexity}, and \ref{assm:continuity} hold, for a given environmental parameter $\mathbf{e}_{\mathsf{tr}}$ and a source location $\ps$, for any perturbation $\boldsymbol{\epsilon}$ that $\|\boldsymbol{\epsilon}\| \leq \frac{\theta }{2 L}$, where $\theta = \min \{\sigma, 1/\xi\}$, the perturbed loss function $\Lcal_{DA}(\mathbf{e}_{\mathsf{tr}} + \boldsymbol{\epsilon}, \ps, \Vec{v})$ has a local minimum in the cube $C_{\rho}(\Vec{v}_0)$, where $\rho = \|\theta \Vec{H}(\Vec{v}_0)^{-1} \Vec{1}]\|_{\infty} \leq \theta / \sqrt{\lambda} $. This implies that
    \begin{multline*}
        \|  \Gamma(\param_{\mathsf{tr}}, \gamma, \hat{\p}_0, \Acal(\mathbf{e}_{\mathsf{tr}}, \ps, .)) -  \Gamma(\param_{\mathsf{tr}}, \gamma, \hat{\p}_0, \Acal(\mathbf{e}_{\mathsf{tr}}+\boldsymbol{\epsilon}, \ps, .)) \| \\ 
    \leq \theta \sqrt{\frac{n_w+n_p}{\lambda}}.
    \end{multline*}

\end{theorem}

\begin{definition} [Positive or Negative Vector]
    We refer to a vector $\Vec{x}$ as ``positive'' (``negative'') if all of its entries are positive (negative), or equivalently, $\mathsf{diag}(\Vec{x})$ is positive (negative) definite.
\end{definition}

\textbf{Outline of the proof}: We show that by strong convexity of the loss function, there exist a positive and a negative gradient vector close to the $\Vec{v}_0$. Then by Lipschitz continuity, we show that under the small perturbation in the environment, the positive (negative) vector will remain positive (negative). Finally, using the Poincar\'e-Miranda theorem \cite{poincare-miranda}, we can prove that the gradient will vanish at a point close to $\Vec{v}_0$, i.e., a local minimum.\par  

\begin{IEEEproof}
We inspect $\mathbf{g}(\kappa)$ element-wise to show that in the training environment, there exists a point with a strictly positive gradient and a point with a strictly negative gradient. Suppose that $0 < \kappa \leq \min \{\sigma, 1 / \xi\}$. We have
\begin{align}\label{eq:g_inequality}
    [\mathbf{g}(\kappa)]_i & = [\mathbf{g}'(0)]_i \kappa + \frac{[\mathbf{g}''(z)]_i} 2 \kappa^2 \nonumber \\
    & \geq [\mathbf{g}'(0)]_i \kappa - \frac{|[\mathbf{g}''(z)]_i|} 2 \kappa^2 \nonumber \\
    & \stackrel{(a)}{\geq} \kappa \left[ [\mathbf{g}'(0)]_i - \frac{|[\mathbf{g}''(z)]_i|} 2 \frac{1}{\xi} \right] \nonumber \\
    & \stackrel{(b)}{\geq} \kappa \left[ [\mathbf{g}'(0)]_i - \frac{1}{2} \right] \stackrel{(c)}{\geq} \frac{ \kappa}{2},
\end{align}
where $(a)$ holds because $0 < \kappa \leq 1/\xi $, $(b)$ holds because of \eqref{eq:xi_bound}, and $(c)$ holds because $\mathbf{g}'(0) = \Vec{1}$. By substituting $\kappa = \theta = \min \{\sigma, 1 / \xi\}$ in \eqref{eq:g_inequality}, we see that the gradient at $\Vec{v}^+ = \Vec{v}_0 + \theta \Vec{H}(\Vec{v}_0)^{-1} \Vec{1}$ is strictly positive and $[G(\Vec{v}^+, \mathbf{e}_{\mathsf{tr}}, \ps)]_i \geq \theta / 2$. Using a similar approach and setting $0 > \kappa \geq \max \{-\sigma, -1/\xi\}$, we can also show the existence of a point with a negative gradient at $\Vec{v}^- = \Vec{v}_0 - \theta \Vec{H}(\Vec{v}_0)^{-1} \Vec{1}$ and $[G(\Vec{v}^-, \mathbf{e}_{\mathsf{tr}}, \ps)]_i \leq - \theta / 2$.

Now we show that $G_{\boldsymbol{\epsilon}} = G(\Vec{v}, \mathbf{e}_{\mathsf{tr}} + \boldsymbol{\epsilon}, \ps)$ also has a negative and a positive vector at $\Vec{v}^- = \Vec{v}_0 - \theta \Vec{H}(\Vec{v}_0)^{-1} \Vec{1}$ and $\Vec{v}^+ = \Vec{v}_0 + \theta \Vec{H}(\Vec{v}_0)^{-1} \Vec{1}$, respectively.
\begin{align}
& \vline \left[G(\Vec{v}^+, \mathbf{e}_{\mathsf{tr}} + \boldsymbol{\epsilon}, \ps)\right]_i - \left[G(\Vec{v}^+, \mathbf{e}_{\mathsf{tr}}, \ps)\right]_i \; \vline \nonumber \\
&\quad \leq \| G(\Vec{v}^+, \mathbf{e}_{\mathsf{tr}} + \boldsymbol{\epsilon}, \ps) - G(\Vec{v}^+, \mathbf{e}_{\mathsf{tr}}, \ps) \| \nonumber \\
&\quad \stackrel{(a)}{\leq} L \|\boldsymbol{\epsilon}\| \stackrel{(b)}{\leq} \theta/2 \leq \; \left[G(\Vec{v}^+, \mathbf{e}_{\mathsf{tr}}, \ps)\right]_i,
\end{align}
where $(a)$ is due to Assumption~\ref{assm:Lipschitz} and $(b)$ is according to the assumption in Theorem \ref{thm:PLSP}. We define $G^+ = G(\Vec{v}^{+}, \mathbf{e}_{\mathsf{tr}}, \ps)$ and $G_{\mathbf{\epsilon}}^+ = G(\Vec{v}^{+}, \mathbf{e}_{\mathsf{tr}} + \boldsymbol{\epsilon}, \ps)$. Since
\begin{align*} 
\left[G_{\mathbf{\epsilon}}^+\right]_i &= \left[G^+\right]_i - \left( \left[G^+\right]_i -  \left[G_{\mathbf{\epsilon}}^+\right]_i \right) \\
& \geq \left[G^+\right]_i - \; \vline \; \left[G^+\right]_i -  \left[G_{\mathbf{\epsilon}}^+\right]_i \; \vline \; \geq 0,
\end{align*}
we deduce that $G(\Vec{v}^{+}, \mathbf{e}_{\mathsf{tr}} + \boldsymbol{\epsilon}, \ps)$ is a positive vector. Similarly, one can show that $G(\Vec{v}^{-}, \mathbf{e}_{\mathsf{tr}} + \boldsymbol{\epsilon}, \ps)$ is a negative vector. Consequently, using the Poincar\'e-Miranda theorem \cite{poincare-miranda}, we conclude that $G_{\boldsymbol{\epsilon}}$ has a root in the $\ell_{\infty}$-neighborhood $C_{\rho}(\Vec{v}_0)$, where 
\begin{align*}
    \rho &= \max_i |[\theta \Vec{H}(\Vec{v}_0)^{-1} \Vec{1}]_i| \\ 
    &= \|\theta \Vec{H}(\Vec{v}_0)^{-1} \Vec{1}\|_{\infty} \leq \theta \sigma_{\max}(\Vec{H}(\Vec{v}_0)^{-1}).
\end{align*} 
where $\sigma_{\max}((\Vec{H}(\Vec{v}_0)^{-1}))$ denotes the largest singular value of $\Vec{H}(\Vec{v}_0)^{-1}$. Because of the $\lambda$-strong convexity of $\Vec{H}(\Vec{v}_0)$, we have that $ \sigma_{\max}((\Vec{H}(\Vec{v}_0)^{-1})) \leq 1/ \sqrt{\lambda}$. Moreover, since $\rho$ is the upper bound on the element-wise perturbation of the vector $\Vec{v}_0 \in \mathbb{R}^{n_w+n_p}$, the norm of the perturbation vector is bounded by $\rho \sqrt{n_w+n_p}$, which in turn yields 
\begin{multline*}
\|  \Gamma(\param_{\mathsf{tr}}, \gamma, \hat{\p}_0, \Acal(\mathbf{e}_{\mathsf{tr}}, \ps, .)) -  \Gamma(\param_{\mathsf{tr}}, \gamma, \hat{\p}_0, \Acal(\mathbf{e}_{\mathsf{tr}}+\boldsymbol{\epsilon}, \ps, .)) \| \\
    \leq \theta \sqrt{\frac{n_w+n_p}{\lambda}}.
\end{multline*}
\end{IEEEproof}

\begin{corollary}
The implication of Theorem \ref{thm:PLSP} is that given a proper forward model (that satisfies the assumptions), equipped with a GB optimization method, the algorithm can be adapted (i.e., incur a bounded error) to a new environment by jointly optimizing over the model parameters $\param$ as well as the input $\p$ during the inference.
\end{corollary}

\section{Empirical Evaluation}
\paragraph*{Setup}We have generated training data corresponding to an environment of depth $200$ m and a constant sound speed of $c = 1500$ m/s and an observation time of $T=2$ s. The receiver is a single hydrophone at depth $120$ m, while the source is at depth $20$ m and at a horizontal range of $610$ m from the receiver. Moreover, the SNR is defined as $\frac{\int_0^T r^2(t) \dt}{B N_0}$, where $B$ is the source pulse bandwidth and $N_0$ is the one-sided noise power spectral density. Also, all results are averaged over $500$ Monte Carlo noise realizations.\par

\begin{figure*}
\centering
\begin{subfigure}[t]{0.4\textwidth}
\centering
    \includegraphics[width= \linewidth]{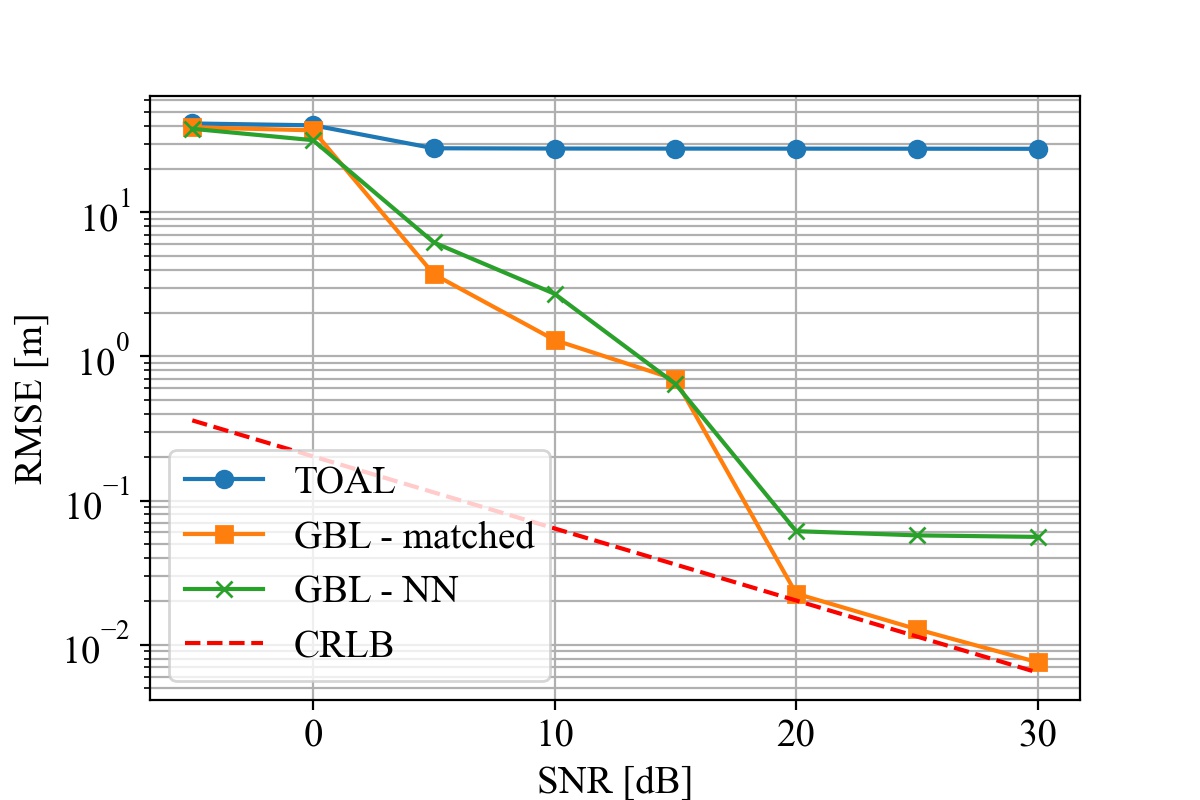}
    \caption{Training depth and the true depth are both $200$ m. The GBL-matched model has full knowledge of the model that created the data, while the GBL-NN model tries to learn the environment using the neural network model in Fig. \ref{fig:modular}.}
    \label{fig:rmse_vs_snr_no_mismatch}
\end{subfigure}
~
\begin{subfigure}[t]{0.4\textwidth}
\centering
    \includegraphics[width= \linewidth]{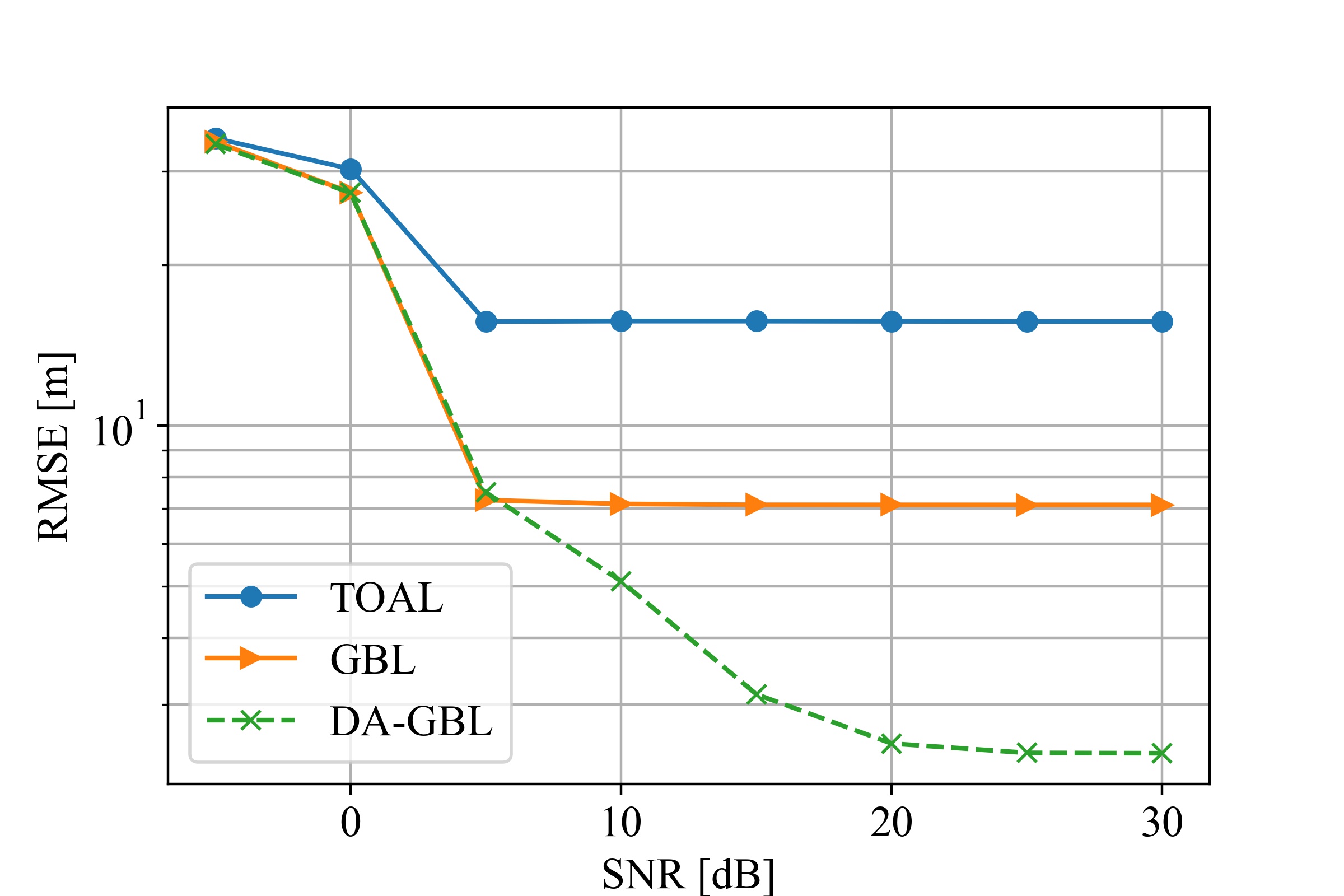}
    \caption{Training depth is $200$ m and the true depth is $198$ m. This result shows how adaptation can improve the localization performance in this mismatched environment.}
    \label{fig:rmse_vs_snr_with_mismatch}
\end{subfigure}
\caption{RMSE performance under different SNR values.}
\label{fig:rmse_vs_snr}
\end{figure*}

\begin{figure*}
\centering
\begin{subfigure}[t]{0.4\textwidth}
\centering
    \includegraphics[width= \linewidth]{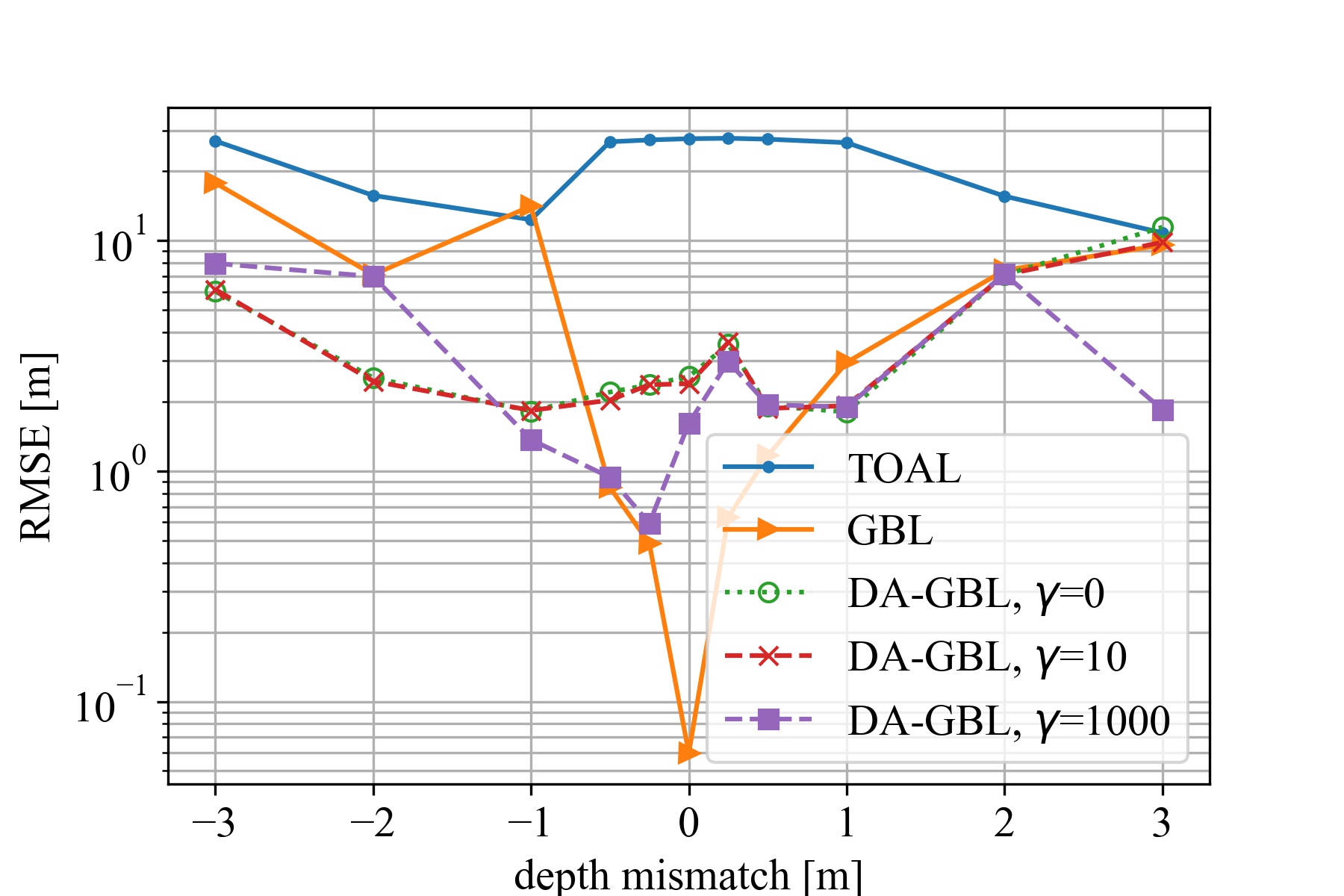}
    \caption{Small mismatches. The adaptation mechanism can overcome small mismatches.}
    \label{fig:small_mismatch}
\end{subfigure}
~
\begin{subfigure}[t]{0.4\textwidth}
\centering
    \includegraphics[width= \linewidth]{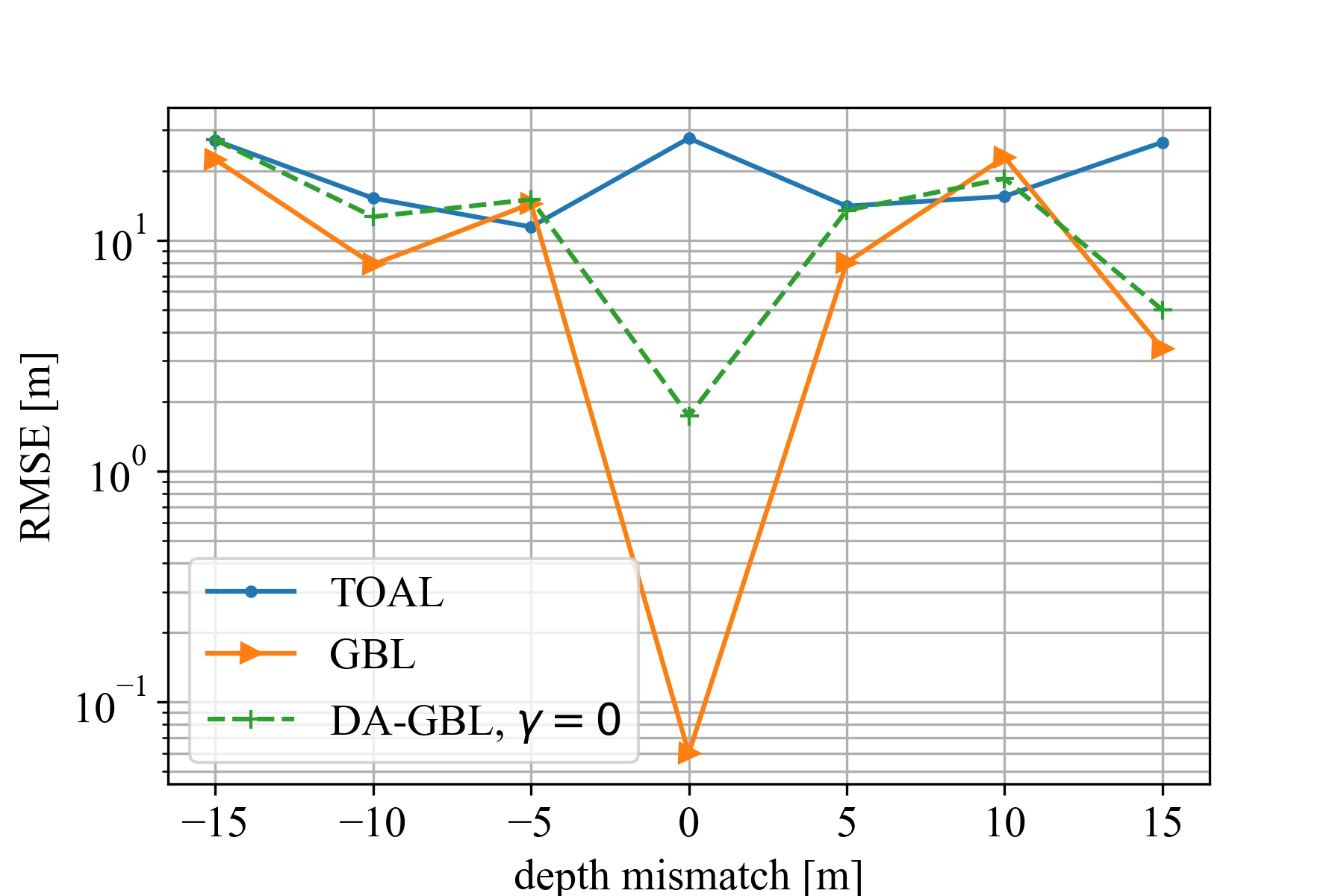}
    \caption{Large mismatches. The adaptation mechanism is not guaranteed to work under large mismatches.}
    \label{fig:large_mismatch}
\end{subfigure}
\caption{RMSE performance under different mismatch values. Training depth is $200$ m and SNR $= 20$ dB.}
\label{fig:different_mismatch}
\end{figure*}

Figures \ref{fig:rmse_vs_snr} and \ref{fig:different_mismatch} provide the simulation results for the $3$-ray environment. According to Fig.~\ref{fig:rmse_vs_snr_no_mismatch}, when there is no mismatch between the training and test environments, the GB approach gets closer to the Cramer-Rao Lower Bound (CRLB) as the SNR increases. Figure \ref{fig:rmse_vs_snr_no_mismatch} shows how a model that has full knowledge of the environment and uses the GBL method can reach CRLB at SNR$=20$ dB. However, after SNR$=20$ dB, the decrease in root mean squared error (RMSE) of the GBL-NN, which has learned the environment from the training data, is negligible. This might be due to the training procedure of the PLN, i.e., the accuracy of the localization is limited by the accuracy of the pre-trained PLN.\par

Furthermore, Fig.~\ref{fig:rmse_vs_snr_with_mismatch} shows that when the test environment has a depth of $198$ m, the ordinary GBL performs poorly due to the $-2$ m mismatch between the training and test environment depths. Nevertheless, the proposed DA-GBL is able to significantly improve the result in this scenario.\par

Figures \ref{fig:small_mismatch} and \ref{fig:large_mismatch} investigate the effects of small and large mismatches respectively. Under large mismatches, the proposed adaptation mechanism will be ineffective, whereas under small mismatches, the proposed method usually reduces the error significantly. Also, Figure \ref{fig:small_mismatch} shows how using a larger $\gamma$ yields a superior performance at smaller mismatches. This is because the algorithm resists changing the pre-trained weights of the model.\par

\section{Conclusion} \label{sec:conclusion}
We have shown that by using a GB optimization along with a pre-trained forward model, we can provide a robust UWA localization algorithm by simultaneous optimization over both the input (location) and the model parameters. Although the theoretical derivations do not guarantee that the model adapts to the new environment with a single unlabeled observation, they show that under small mismatches, the algorithm is robust. Furthermore, we have removed the need for peak labeling by enforcing the governing physical laws into the architecture of the network, which enables the learning of the peak labels implicitly. While this architecture can be used straightforwardly in similar problems such as room acoustics, to exploit it in ocean acoustics, one needs to tailor it to the non-constant sound speed nature of the environment.\par



\end{document}